\documentclass[journal=jacsat,manuscript=article,layout=twocolumn]{achemso}
\let\oldmaketitle\maketitle
\let\maketitle\relax

\usepackage{chemformula} 
\usepackage[T1]{fontenc} 
\usepackage{soul}

\newcommand{\rev}[1]{\textcolor{black}{#1}}

\author{Rodrique G. M. Badr}
\email{rbadr@uni-mainz.de}
\affiliation{Institut f\"ur Physik, Johannes Gutenberg-Universit\"at Mainz, Staudingerweg
	7-9, D-55099 Mainz, Germany}

\author{Lukas Hauer}
\affiliation{Max Planck Institute for Polymer Research, Ackermannweg 10, 55128 Mainz, Germany}

\author{Doris Vollmer}
\affiliation{Max Planck Institute for Polymer Research, Ackermannweg 10, 55128 Mainz, Germany}

\author{Friederike Schmid}
\email{friederike.schmid@uni-mainz.de}
\affiliation{Institut f\"ur Physik, Johannes Gutenberg-Universit\"at Mainz, Staudingerweg
	7-9, D-55099 Mainz, Germany}

\title[Dynamics of Droplets Moving on Lubricated Polymer Brushes]
  {Dynamics of Droplets Moving on Lubricated Polymer Brushes}


\begin{document}



 \twocolumn[
 \begin{@twocolumnfalse}
 \oldmaketitle

\begin{abstract}
Understanding the dynamics of drops on polymer-coated surfaces is crucial for optimizing applications such as self-cleaning materials or microfluidic devices. While the static and dynamic properties of deposited drops have been well characterised, a microscopic understanding of the underlying dynamics is missing. In particular, it is unclear how drop dynamics depends on the amount of uncrosslinked chains in the brush, because experimental techniques fail to quantify those. Here we use coarse-grained simulations to study droplets moving on a lubricated polymer brush substrate under the influence of an external body force. The simulation model is based on the many body dissipative particle dynamics (mDPD) method and designed to mimic a system of water droplets on polydimethylsiloxane (PDMS) brushes with chemically identical PDMS lubricant. In agreement with experiments, we find a sublinear power law dependence between the external force $F$ and the droplet velocity $v$, $F \propto v^\alpha$ with $\alpha <1$; however, the exponents differ ($\alpha \sim 0.6-0.7$ in simulations versus $\alpha \sim 0.25$ in experiments). With increasing velocity, the droplets elongate and the receding contact angle decreases, whereas the advancing contact angle remains roughly constant. Analyzing the flow profiles inside the droplet reveals that the droplets do not slide, but roll, with vanishing slip at the substrate surface. Surprisingly, adding lubricant has very little effect on the effective friction force between the droplet and the
substrate, even though it has a pronounced effect on the size and structure of the wetting ridge, especially above the cloaking transition. 

\end{abstract}

\bigskip

\end{@twocolumnfalse}
]



\section{Introduction}

Droplets are ubiquitous in nature and technology. Understanding the wetting dynamics of liquid droplets enables improved use of droplets for various applications\citep{brutin2018recent,wang2022wetting,bertrand2002wetting,wong2011bioinspired,lafuma2003superhydrophobic,schellenberger2015direct,wong2021super}, including self-cleaning\citep{blossey2003self,nakajima2000transparent,parkin2005self}, spray coating\citep{tejero-martin2019beyond}, efficient application of pesticides\citep{bergeron2000controlling,gaume2002function}, microfluidics\cite{choi2012digital}, heat transfer\citep{cho2017nanoengineered}, or drag reduction\citep{rothsthein2010slip,lee2016superhydrophobic,park2021superhydrophobic}.

Coating a solid substrate by polymer brushes or gels provides a versatile way to tune the static and dynamic properties of deposited drops\citep{xu2008directing, liu2010switching, liu2011photoregulated, karpitschka2015droplets,karpitschka2016liquid, lee2016autophobic,okada2020thermo, khatir2023molecularly,kajouri2023antidurotaxis,kajouri2023unidirectional}.
In particular, the softness, type, and thickness of the coating govern the friction a drop experiences when sliding or rolling over a surface. The dynamic wetting properties of drops on soft surfaces result from the interplay between interfacial tension, contact line tension, substrate elasticity, and dissipation within the substrate and drop \citep{weijs2011origin, weijs2014capillarity,andreotti2016soft, andreotti2020statics, daniel2020quantifying,peschka2018variational,giacomelli2023droplet,cai2021fluid,hauer2023phase}. This intricate puzzle of coexisting effects makes understanding, predicting, and tuning the friction of drops on soft surfaces challenging.

Polydimethylsiloxane brushes form a class of promising soft coatings for applications \citep{wang2016covalently,chen2023omniphobic}.
A PDMS brush is composed of PDMS chains grafted by one end to a surface. The resulting viscoelastic layer has a thickness of a few nm. PDMS chains are very flexible, having a persistence length of a few monomers \citep{zhao2021macroscopic}. In the brush, the chains keep their flexibility as opposed to PDMS gels where the chains are crosslinked, resulting in an elastic solid. Owing to the flexibility of the brush chains, water drops on PDMS can show a contact angle hysteresis of less than $10^\circ$\citep{wooh2016silicone}. Often PDMS brushes are swollen by PDMS chains which are not grafted to the surface.
These mobile chains may provide additional lubrication\citep{teisala2020grafting}.

One characteristic of wetting on soft surfaces is the deformation of the surface surrounding a deposited drop. The vertical component of the interfacial tension pulls the substrate up. A wetting ridge forms \cite{shanahan1994anomalous,shanahan1995viscoelastic, long1996static,roy2020droplets}. 
The shape and height of the ridge result from the competition between the surface tension of the droplet applying a vertical stress on the substrate and the substrate elasticity\citep{bico2018elastocapillarity}. The wetting ridge 
plays a crucial role in the dynamics of sliding droplets, with the main effect being viscoelastic energy dissipation inside the ridge\citep{karpitschka2015droplets,zhao2018geometrical,sadullah2018drop,peschka2018variational,giacomelli2023droplet}. This so-termed contact line friction contributes to the energy dissipation during the motion of the droplet.

In addition, mobile chains can remain in the PDMS film. The poor solubility of silicone oil in most solvents requires severe rinsing of the film to remove all mobile chains. Since the amount of remaining mobile chains is hard to quantify experimentally, its contribution to dynamic wetting is still unclear \citep{teisala2020grafting,smith2013droplet,schellenberger2015direct}.

This is further complicated by the fact that mobile chains can cloak a drop. The cloaking of water by silicone oil on lubricated surfaces influences both static and dynamic experiments\citep{daniel2017oleoplaning,kreder2018film}. It has the effect of dynamically reducing the surface tension of the drop \citep{hourlier2018extraction,naga2021water}, and contributes to the depletion of lubricant for drops rolling off on surfaces as the cloaking material leaves the surface with the drop. 

Despite the interesting physics and applications of wetting on PDMS brushes, recent experimental work on dynamic wetting on PDMS has largely focused on gels \citep{jeon2023moving,hauer2023phase}. This is partly due to the fact that polymer brushes typically have thicknesses in the order of nanometers, which makes it challenging to probe the microscopic details of the phenomena in question.
Numerical methods can provide valuable insights. In particular, numerical methods allow us to look into details such as the flow \citep{thampi2013liquid} and dissipation within the droplet \citep{servantie2008statics}, the structure of the wetting ridge on polymer brushes \citep{leonforte2011statics,badr2022cloaking}, and the structure of the cloaking layer \citep{badr2022cloaking}. 

To date and to our best knowledge, simulations of droplets on polymer brushes have been limited to the study of static properties \citep{leonforte2011statics, cao2015polymeric,mensink2019wetting,mensink2021role,badr2022cloaking}, or dynamic properties for brushes that are swollen by the drop \citep{greve2023stick,kajouri2023antidurotaxis,kajouri2023unidirectional,kap2023nonequilibrium} or serve as coatings in nanochannels\citep{speyer2017droplet, pastorino2021liquid,leong2021dynamics}. The dynamics of drops rolling on PDMS brushes remains an open question. In the present study, we use coarse-grained molecular dynamics simulations to address this problem. One question of integral importance is the dependence of drop dynamics on degree of lubrication. In order to study this, we first investigate the relation between friction forces and velocity, which provides a global measure of energy dissipation. This is mostly done by simulations, but we also present corresponding experimental studies. The dissipation stems from multiple sources, such as the slip between the droplet and the brush, viscoelastic dissipation in the ridge, the flow within the droplet... To better understand the active players, we then use simulations to quantify the response of the brush and lubricant, and the flow within the droplet. Another standing question is that of the transport and depletion of lubricant. The cloak is expected to enhance the lubricant transport. However, it is not clear whether the cloak is maintained for moving droplets, and if so how the velocity of the droplet affects the cloak. To better understand this, we characterize the cloaking layer for droplets moving with different velocities.


\section{Model and Methods}
\label{sec:sim}


\subsection{Simulations}

\subsubsection{Simulation Model}

We consider a coarse-grained model system containing a polymer brush, free lubricant chains, and a droplet made of liquid particles, in coexistence with a vapor phase. Polymers are chains of beads connected by springs with the spring potential $U_{bond}=k(r_{ij}-r_0)^2$. Liquid molecules are modeled as single isolated beads. To model the non-bonded potential and the coupling to a heat bath at a given temperature $T$, we use the Many-body Dissipative Particle Dynamics (MDPD) coarse-grained model and thermostat. The DPD thermostat has the advantage that the dissipative and random forces are pairwise interactions and momentum conserving allowing for hydrodynamic phenomena \citep{hoogerbrugge1992simulating,espanol1995hydrodynamics,marsh1997static}, while the multi-body force element allows for modeling the coexistence of two phases\citep{pagonabarraga2002mdpd, trofimov2002mdpd, warren2003vapor}. The forces take the following form:


\begin{align}
		&F_{ij}=F_{ij}^C+F_{ij}^D+F_{ij}^R\\
		&F_{ij}^C=A_{ij}w^C(r_{ij})\hat{r}_{ij}+B_{ij}(\bar{\rho}_i+\bar{\rho}_j)\tilde{w}^C(r_{ij})\hat{r}_{ij}\\
		&F_{ij}^D=-\zeta w^C(r_{ij})^2(\hat{r}_{ij}.\vec{v}_{ij})\hat{r}_{ij}\\
		&F_{ij}^R=\sqrt{2\zeta k_B T}w^C(r_{ij})\theta_{ij}\hat{r}_{ij}\\
		&w^C(r_{ij})=\begin{cases}
			\bigg(1-\frac{r_{ij}}{r_c}\bigg)~~&r_{ij}\leq r_c\\
			0~~&r_{ij}>r_c
		\end{cases} \\
		&\tilde{w}^C(r_{ij})=\begin{cases}
			\bigg(1-\frac{r_{ij}}{r_d}\bigg)~~&r_{ij}\leq r_d\\
			0~~&r_{ij}>r_d
		\end{cases}\\
		&\bar{\rho}_i=\sum_{j\neq i}\frac{15}{2\pi r_d^3}\tilde{w}^C(r_{ij})^2
\end{align}

\noindent In the above equations, $F_{ij}^C$ is the conservative force contribution where $A_{ij}<0$ is the strength of the attractive part, and $B_{ij}>0$ is the strength of the density dependent repulsion. $B_{ij}$ must have the same value for all pairs of particles for the forces to be conservative as shown by the no-go theorem of MDPD \citep{warren2013no}. We also have $\vec{r}_{ij}=\vec{r}_i-\vec{r}_j$, and $\hat{r}_{ij}=\vec{r}_{ij}/r_{ij}$. $F_{ij}^D$ and $F_{ij}^R$ are the dissipative and random force contributions respectively, where $\zeta$ is the drag coefficient, $\vec{v}_{ij}=\vec{v}_i-\vec{v}_j$, $k_B$ and $T$ are Boltzmann’s constant and the temperature respectively, and $\theta_{ij}$ is an uncorrelated Gaussian distributed random variable with zero mean and unit variance. $w_C$ and $\tilde{w}_C$ are weight functions, $\tilde{\rho}_i$ is a weighted density, and finally $r_c$ and $r_d$ are cutoff radii which set the range of the forces. The reason for introducing two cutoff radii is that the range of the density-dependent repulsion must be smaller than that of the attraction $r_d<r_c$ (with $A_{ij} < 0$ and $B_{ij} > 0$) in order to obtain liquid-vapor coexistence \citep{warren2003vapor}.

The polymer brush consists of end-grafted chains at varying grafting densities. The chains are grafted to a purely repulsive surface modeled using the Weeks-Chandler-Anderson (WCA) potential \citep{weeks1971role}. The lubricant is modeled as free chains. The free and grafted chains are all taken to be of the same species and therefore have the same interaction parameters among each other and with the liquid particles.

The simulations are performed in the NVT ensemble, using periodic boundary conditions in all directions. The unit of energy is set by $k_BT=1$, the unit of length by the cutoff distance of the DPD attraction $r_c=1$, and the mass unit by $m=1$ for all species. The unit of time can then be defined as $\tau=\sqrt{\frac{m r_c^2}{k_BT}}$. In the following, all quantities are given in these units. Other fixed parameters include $r_d=0.8;~\zeta=4.5;~B_{ij}\equiv B=40;~dt=10^{-3}$. For the grafting surface WCA potential we choose $\sigma_{WCA}=1,\epsilon=1$. With this choice of parameters and the thickness of the brush, the grafting surface does not directly influence the wetting behavior. The spring constant $k=20$ and equilibrium extension $r_0=1$ are chosen for the bond potential. The resulting bond length is $a\approx1.09$. All simulations are performed in the absence of any gravitational forces.

To produce a system mimicking water on PDMS, we choose the interaction parameters as in earlier work \cite{badr2022cloaking}, such that the surface tensions between the different phases reproduce the relevant contact angles, i.e. contact angle of water on bulk PDMS materials. To this end, we choose the polymer-polymer cohesion as $A_{pp}=-28$, the liquid-liquid cohesion as $A_{ll}=-50$, and the polymer-liquid cohesion as $A_{pl}=-21$. With this choice, we obtain values for the liquid-vapor interfacial tension $\gamma_{w}\approx3.2\, k_B/r_c^2$, the lubricant-vapor tension $\gamma_{o}\approx0.9\,k_BT/r_c^2$, the lubricant-liquid tension $\gamma_{ow}\approx1.4\, k_BT/r_c^2$. This gives an expected Young contact angle\citep{young1805iii} of $\theta_Y\approx100$ and a positive spreading parameter of lubricant on the liquid, $S_{ow}=\gamma_{w}-\gamma_{ow}-\gamma_{o}\approx0.9$. The dynamic viscosity of the liquid has been determined by Poiseuille flow simulations (see SI), giving $\eta = 6.4\, k_B T r_c^{-3} \tau$. The equilibrium densities of the liquid and the lubricant are respectively $\rho_w \approx 4$ and $\rho_o \approx 2.9$.
 
The simulations are conducted using the HOOMD-Blue simulation package \citep{anderson2020hoomd,phillips2011pseudo} version 2.9.7. All snapshot visualizations are made with the OVITO visualization package \citep{stukowski2009visualization}.


\subsubsection{System preparation}

The brush is composed of $n_B$ chains of length $N_B=50$ with the first monomer of each chain fixed on the grafting surface following a regular square lattice pattern. We choose the lattice constant $d=2\,r_c$, corresponding to a grafting density $\sigma=0.25\,r_c^{-2}$. We use $n_B = 100\times100$ chains resulting in typical box sizes of $200 \times 200 \times 300 \: r_c^3$.

In addition, the system may contain $n_o$ lubricant polymers of length $N_o=5$. We characterize the amount of lubricant present in terms of the number fraction of lubricant monomers to brush monomers,
\begin{equation}
	\Phi=\frac{n_oN_o}{n_oN_o+n_BN_B}.
\end{equation}
The brush is first equilibrated without any lubricant for $8\times10^5$ simulation steps. We separately prepare a film of lubricant of length $N_o=5$ at the equilibrium melt density $\rho_o\approx2.9\, r_c^{-3}$. The lubricant is then added to the system by placing the film in contact with the brush and letting the lubricant infuse the brush. Equilibrium is reached after a maximum of $64\times10^5$ steps. 

To prepare the spherical droplets we take a pendant droplet and place it in contact with the brush, and let it equilibrate in the absence of gravity. The number of liquid particles was $n_w =766\times10^3$ in all simulations. The systems are then left to equilibrate until no more conformational changes are observed. The systems with the droplet need at least $10^7$ steps to reach equilibrium. This results in the equilibrium configurations described in our previous publication \citep{badr2022cloaking}, \rev{where we find that a cloaking transition exists, with the cloaking setting in at $\Phi \approx 0.4$. Therefore,} some droplets will be fully cloaked at equilibrium, depending on the fraction of free chains present in the system. 

Finally, to induce the motion of the droplet, we apply a constant force of magnitude $F$ in the positive x-direction. The force is applied to every liquid particle and only to the liquid particles. This mimics the experimental situation where the force is transmitted to the droplet by a cantilever (see below). The simulations with the force are then run for at least $24\times10^5$ steps until a steady state with constant droplet velocity is reached. All measurements are made during a consequent run of $48\times10^5$ steps where 150 simulation snapshots are obtained. For each set of parameters, 5 independent simulations are performed. \rev{The independent simulations are obtained by equilibrating 5 different polymer brushes from the same initial condition, but with different initial seeds for the random number generator. Following that, every stage of the system preparation is again conducted with fresh set of seeds for the random number generator, resulting in minimally correlated simulations.} The results from those simulations are averaged to obtain mean values $\bar{x}$ and the error bars in the plots correspond to standard errors $\alpha_x=\sqrt{ \frac{ \sum_{i=1}^N (x_i-\bar{x})^2 }{ N(N-1) } }$ with $N=5$ is the number of independent values.

\subsection{Experimental Setup}

\rev{In the experiments we measure the friction force experienced by a droplet moving at a constant velocity. This is with the aim of comparing the results with the corresponding calculations from simulations. While a direct comparison of the values for forces and velocities is not possible due to the difference in units, we can still compare trends, laws, etc\ldots}

Glass slides ($24\times 60~\mathrm{mm^2}$, $170\pm 5~\mathrm{\mu m}$ thickness, Menzel-Gläser) were coated with PDMS pseudo brushes, using the ''grafting-from'' \citep{zhao2020non,hegner2023fluorine} and the ''grafting-to'' \citep{teisala2020grafting} approaches. In short, ''grafting-from'' implies the polymerization of vaporous monomers from grafting sites on the surface while prepolymerized chains are ''grafted-to'' grafting sites. For grafting sites, hydroxyl groups are formed by $O_2$-plasma activation ($300~\mathrm{W}$, $10~\mathrm{min}$) of the glass surfaces. For ''grafting-from'', we placed the plasma-activated glass slides in a desiccator ($20^\circ$C, $40\%$ humidity ) together with 1 ml dichlorotetramethyl-disiloxane (Sigma-Aldrich). After 10 min the samples were removed from the desiccator. The number or percentage of free PDMS chains can not be quantified experimentally because the amount is insufficient even for state of the art analytical techniques. For ''grafting-to'', tetramethyl-terminated PDMS oil ($6~\mathrm{kDa}$ and $N_b\approx80$, Alfa Aesar) was drop cast on the activated glass slides. Atmospheric or surface-bound water can break Si-O bonds, allowing PDMS chains to bind to the surface hydroxyl groups. The samples were equilibrated overnight. Excess oil was removed by sonicating the samples for 30 min in toluene.

Sliding forces were measured with a cantilever set-up, mounted on a confocal microscope (Leica, SP8) \citep{naga2021water}. The coated glass slides are mounted on a motorized stage, above the microscope lens. An approx. $100~\mathrm{mm}$ long metal blade is vertically placed over the coated glass slides so that the upper end is fixed while the lower end hangs freely several microns above the surface. $10~\mathrm{\mu l}$ large drops are placed on the coated glass slide and moved at constant speeds against the metal blade. This results in a displacement of the blade that we track with the microscope. The blade displacement is linearly related to a force by the spring constant. We determine the spring constant with a \rev{natural frequency analysis \citep{naga2021capillary}} to be around $214~\mathrm{mN/m}$, yielding a measurable force resolution of approx. $5~\mathrm{\mu N}$. In steady-state, the sliding forces of the droplet balance with the spring force of the displaced blade. 


\section{Results and Discussion}


\subsection{Friction Force and Dissipation}

\begin{figure}[!t]
	\begin{centering}
			\includegraphics[width=8cm]{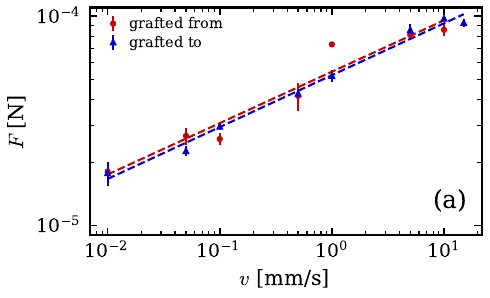}
			\\
		\includegraphics[width=8cm]{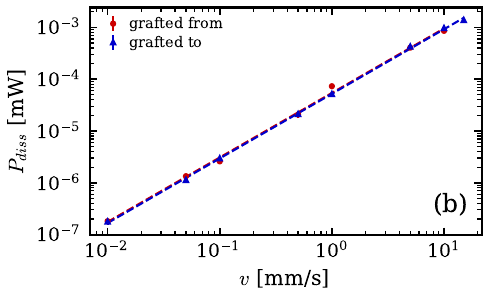}
	\end{centering}
	\caption{ Experimental measurements for two methods for synthesizing PDMS brushes, chemical vapor deposition (grafting from) and drop casting (grafting to). (a) Steady-state force exerted on the cantilever versus velocity $v$ of the stage, on logarithmic axes. The dashed lines are power-law fits $F\propto v^\alpha$ with $\alpha_{from}\approx 0.246\pm0.028$ and $\alpha_{to}\approx 0.248\pm0.011$. (b) Steady-state power dissipation defined as $P_{diss}=F v$ versus the velocity $v$ of the stage, on logarithmic axes. The dashed lines are power-law fits $P_{diss}\propto v^\beta$, with $\beta_{from}\approx 1.246\pm0.028$ and $\beta_{to}\approx 1.248\pm0.011$.}
	\label{fig:expt_results}
\end{figure}

We first investigate the relation between the energy dissipation and the velocity of the droplet relative to the substrate. Experimentally, this is measured by imposing a constant substrate velocity while keeping the droplet in place by a cantilever, and measuring the force on the cantilever \citep{naga2021water} (see Section ''Experimental Setup'' above). The measured force corresponds to the friction force felt by the droplet. Once the force is measured, the power dissipated is calculated as \rev{$P_{diss}=F\cdot v$}. In the experiments, the force was measured for different velocities for droplets deposited on PDMS brushes synthesized using two different methods as described earlier: ''grafting from'' and ''grafting to''. The ''grafting to'' samples were washed after synthesis to remove free chains, while the ''grafting from'' samples were not washed. The results from the experiments are shown in Figure \ref{fig:expt_results}. The force measurements on both samples seem to coincide and show very little effect of the synthesis method. One factor that would influence the friction force is the presence of free chains which would act as lubricant. Since the ''grafting to'' samples were washed, no lubricant should be present \citep{teisala2020grafting}. The experimental measurements then imply: either the ''grafting from'' method does not leave residual free chains, the amount of free chains in PDMS polymer brushes was insufficient to reduce drop friction, or free chains hardly lower friction. The latter would mean that the addition of a small amount of free chains will only swell the brush and not make the interface more liquid-like, hence not lowering drop friction. On the other hand, it has been demonstrated in Reference \citenum{teisala2020grafting} that free chains in a brush reduce the contact angle hysteresis, which is directly proportional to the friction force. Since it is difficult to control the amount of free chains in PDMS brushes experimentally, we turn to the simulations to investigate the effect of free chains on drop friction in a controlled fashion.

\begin{figure}[!htb]
	\begin{centering}
		\includegraphics[width=8cm]{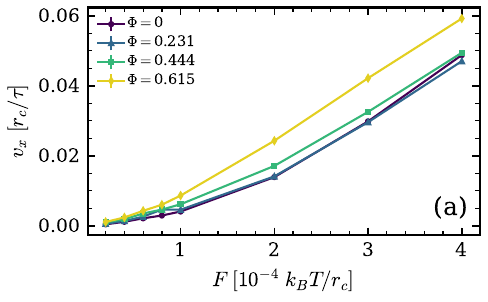}
		\includegraphics[width=8cm]{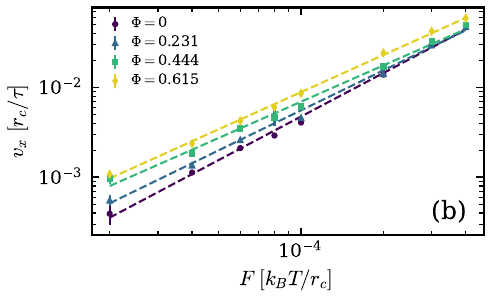}
	\end{centering}
	\caption{(a) Steady-state velocity $v_x$ versus the applied force $F$, on linear axes. (b) Steady-state velocity $v_x$ versus the applied force $F$, on logarithmic axes. The dashed lines show a fit to a power-law behavior $v_x\propto F^{1/\alpha_{sim}}$, with $\alpha_{sim}=0.621\pm0.015; 0.675\pm0.020;0.744\pm0.025;0.727\pm0.013$ for $\Phi=0 ;0.231; 0.444; 0.615$ respectively. $\Phi\geq 0.444$ corresponds to cloaked droplets.}
	\label{fig:sim_FvsV}
\end{figure}

In the simulations, we apply a constant force $F$ to each of the liquid particles and measure the ensuing steady-state velocity $v_x$. The velocity $v_x$ is taken as the mean x-component of the velocity of all liquid particles. The simulations were performed at varying fractions of free chains $\Phi=0;\, 0.231;\, 0.444;\, 0.615$, giving brushes that are partly swollen by free chains, but not yet fully saturated (the saturation point is $\Phi \approx 0.77$)\cite{badr2022cloaking}. For each fraction $\Phi$, forces ranging from $F=2\times 10^{-5} k_BT/r_c$ to $F=4\times 10^{-4}k_BT/r_c$ are applied on the liquid particles. For forces lower than $F=2\times 10^{-5} k_BT/r_c$ the droplet does not seem to move. For forces higher than $F=4\times 10^{-4}k_BT/r_c$ the droplet smears on the brush and forms a ''rivulet'' across the periodic boundary. Figure \ref{fig:sim_FvsV} shows the steady-state velocity $v_x$ versus the force $F$ in linear (a) and logarithmic (b) representation. Looking at the trend lines, the velocity measurements for the lower free chain fractions $\Phi=0;\, 0.231$ essentially coincide, whereas the velocity at the same force increases for $\Phi=0.444$ and again for $\Phi=0.615$, implying that the free chains indeed reduce the friction between the droplet and the substrate. Interestingly, this qualitative change of behavior coincides with the cloaking transition, which was located at $\Phi \sim 0.4$ in this system\cite{badr2022cloaking}: $\Phi=0.444$ is just above the cloaking transition, and the droplets are not cloaked for the lower fractions. The trends in Figure \ref{fig:sim_FvsV}~b) suggest a power law $v_x\propto F^{\alpha_{sim}}$ with exponents ranging between $\alpha_{sim} \sim 0.6$ for bare droplets on dry brushes to $\alpha_{sim} \sim 0.7$ for cloaked droplets on lubricated brushes.

\begin{figure}[t]
	\begin{centering}
		\includegraphics[width=8cm]{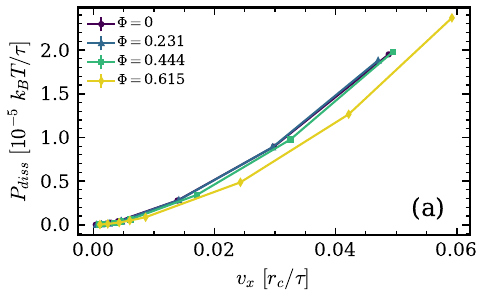}
		\includegraphics[width=8cm]{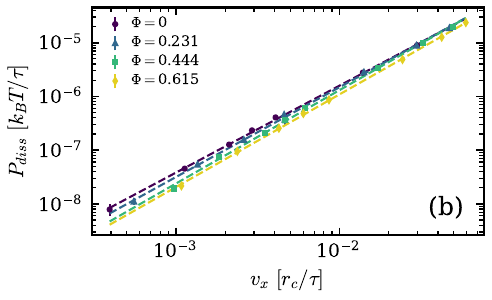}
	\end{centering}
	\caption{(a) Steady-state power dissipation defined as $P_{diss}=F v_x$ versus the steady-state velocity $v_x$, on linear axes. (b) Steady-state power dissipation defined as $P_{diss}=F v_x$ versus the steady-state velocity $v_x$, on logarithmic axes. The dashed lines show fits to a power-law behavior $P_{diss}\propto v_x^{\beta_{sim}}$, with exponents $\beta_{sim}=1+\alpha_{sim}=1.618\pm0.015;1.671\pm0.020;1.739\pm0.025;1.726\pm0.013$ for $\Phi=0;0.231;0.444;0.615$ respectively. $\Phi\geq 0.444$ corresponds to cloaked droplets.}
	\label{fig:sim_PdissvsV}
\end{figure}

Another way of looking at the data is to calculate the power dissipated by the droplet. This can be done like in the experiments using \rev{$P_{diss}=F\cdot v_x$}. Figure \ref{fig:sim_PdissvsV} shows the steady-state power dissipation $P_{diss}$ versus the steady-state velocity $v_x$ with linear axes in (a) and logarithmic in (b). This graph demonstrates again that the dissipated power is lower for higher lubrication (beyond the cloaking transition). The logarithmic plot also suggests a power law relation $P_{diss}\propto v_x^{\beta_{sim}}$ with exponents $\beta_{sim}=1+\alpha_{sim}$ in the range of $\beta \sim 1.6$ for droplets on dry brushes to $\beta \sim 1.7$ for cloaked droplets. 

The power law exponents in experiment and simulation are not the same. We will discuss this further below. One interesting aspect is that the presence of a small amount of free chains has no significant effect on the friction force acting on the droplet. This suggests that the quantity of free chains in the ''grafting from'' method may not be large enough to lubricate the motion of the droplet. Beyond the cloaking transition, the friction between the droplets and the substrate is somewhat reduced. However, even this effect is comparatively small. \rev{As mentioned above, the total friction has contributions from multiple sources. Below, we quantify the response of the substrate and the flow within the droplet to better understand the small dependence of the friction force on lubrication.}


\subsection{Droplet Shape and Contact Angles}
\label{sec:dropletShape}

\begin{figure}[!tb]
	\begin{centering}
		\includegraphics[width=8cm]{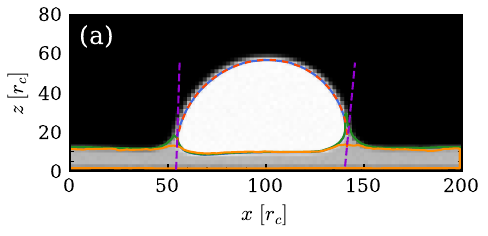}
		\includegraphics[width=8cm]{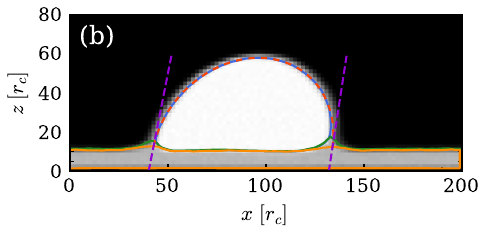}
		\includegraphics[width=8cm]{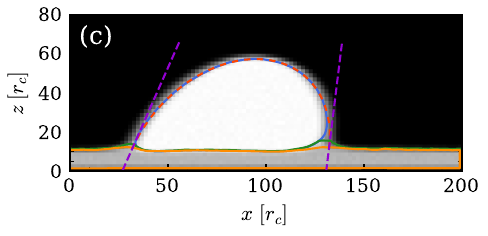}
		\includegraphics[width=8cm]{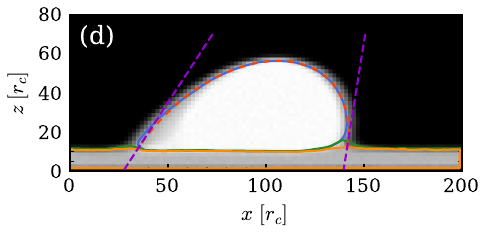}
	\end{centering}
	\caption{Density map showing contours of the brush (orange), the free chains (green), the droplet (blue), along with a fit to an ellipse for the droplet (red dashed line). The fit was then used to find the tangents to the droplet where it intersects the unperturbed height of the brush (purple dashed lines). (a) $F=8\times10^{-5}$, (b) $F=20\times10^{-5}$, (c) $F=30\times10^{-5}$, (d) $F=40\times10^{-5}$. For all figures $\Phi=0.615$. $\Phi\geq 0.444$ corresponds to cloaked droplets.}
	\label{fig:contours}
\end{figure}

As the droplet slides, the forces acting on it cause it to deform, and at steady-state the shape of the droplet can be approximated by an ellipse. In addition, the advancing and receding contact angles differ from each other due to the adhesion of the droplet to the substrate.
In order to characterize these effects, we take a section of the droplet of thickness $2r_c$ along a plane parallel to the direction of motion and analyze the density map in that plane. Examples of such density maps are shown in Figure \ref{fig:contours}. From the density maps we delimit the different interfaces as density isosurfaces. The droplet surface (blue lines in Figure \ref{fig:contours}) corresponds to the water density isosurface $\rho_w^\text{iso} \equiv 2 r_c^{-3}$, and the full substrate surface (green lines) to the polymer density isosurface $\rho_o^\text{iso}\equiv 1.5 r_c^{-3}$ (brush and free chains). These isosurfaces (or density contours) were set at densities corresponding to roughly half the bulk density of water and polymer. The brush surface (orange lines) is taken to be the brush monomer isosurface $\rho_b^\text{iso} \equiv \rho_b^\text{mean}/2$, where $ \rho_b^\text{mean}$ is the mean brush density and depends on the level of lubrication. Having determined the droplet surface, we can now fit the droplet-vapor part to a tilted ellipse. This is achieved by optimizing the coordinates of the foci $F_1$ and $F_2$ such that they best obey the relation $PF_1 + PF_2 = 2A$, with $P$ a point on the droplet-vapor part of the contour, and $A$ the semi-major axis of the ellipse. The parameters $F_1$, $F_2$, and $A$ fully define the ellipse, and we can calculate contact angles at the advancing and receding ends. For this, we find the intersection of the ellipse with the horizontal line at the level of the unperturbed substrate and calculate the angle between the tangent to the ellipse at that point and the horizontal line. With this approach, we measure apparent, macroscopic, contact angles. In Figure \ref{fig:contours}, the ellipse and both the receding and advancing tangents are shown in red and purple dashed lines respectively. The ellipse fits the droplet quite well for all velocities at the receding end. However, right at the advancing front, the fit deviates significantly from the contour of the droplet. Thus the values of macroscopic advancing contact angles differ from the microscopic values. Unfortunately, it is difficult to unambiguously define microscopic contact angles in the presence of a wetting ridge, especially if the free chains spread on the droplet. Therefore, we limit our analysis to macroscopic contact angles.

\begin{figure}[!tb]
	\begin{centering}
		\includegraphics[width=8cm]{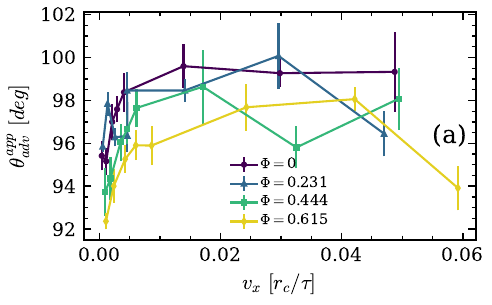}
		\includegraphics[width=8cm]{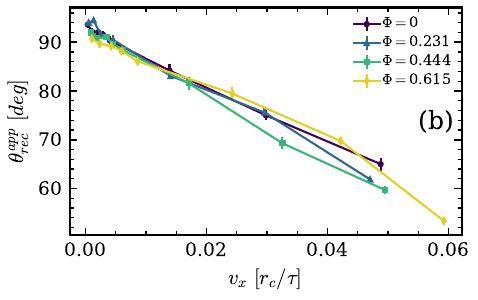}
		\includegraphics[width=8cm]{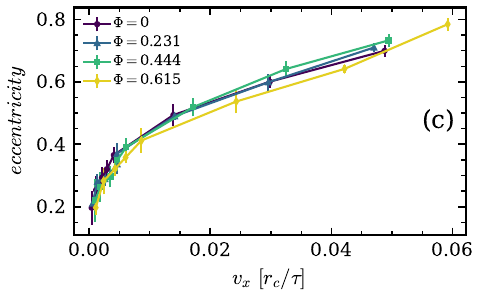}
	\end{centering}
	\caption{(a) Advancing contact angle $\theta_{adv}$ versus droplet velocity. (b) Receding contact angle $\theta_{rec}$ versus droplet velocity showing a roughly linear decrease. (c) Eccentricity of the ellipse that best fits the droplet shape versus the droplet velocity.}
	\label{fig:anglesvsV}
\end{figure}

Figure \ref{fig:anglesvsV}~a) shows the apparent advancing contact angle $\theta_{adv}^{app}$ versus velocity for different fractions of free chains. The shape of the curves is similar to that found in earlier experiments on expanding water drops on PDMS surfaces\cite{william2020adaptive}: After an initial increase, the advancing contact angle quickly saturates, and it overall varies by less than ten degrees. In contrast, the receding contact angle $\theta_{adv}^{rec}$, shown in Figure \ref{fig:anglesvsV}~b), decreases seemingly linearly as the velocity increases and drops to roughly 2/3 of the force-free value at the highest force. Along with this decrease of $\theta_{adv}^{rec}$, the eccentricity of the ellipse strongly increases with velocity (Figure \ref{fig:anglesvsV}~c). However, all three quantities, i.e., the advancing and receding contact angles and the eccentricity, are not significantly affected by the presence of lubricant. This is already observed in equilibrium droplets: The equilibrium contact angles for different lubricant fractions are shown in Figure \ref{fig:equilAngles_Ridge} in the
Supplementary Information. For all lubricant fractions $\Phi$ we find an apparent angle $\theta_{app} \sim 92-95$.


\subsection{The Wetting Ridge}

\begin{figure*}[!t]
	\begin{centering}
		\includegraphics[width=8cm]{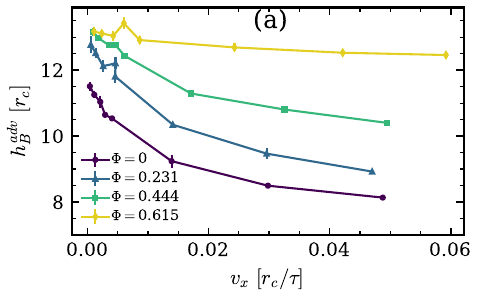}
		\includegraphics[width=8cm]{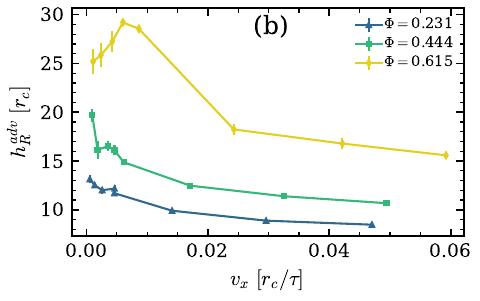}
		\includegraphics[width=8cm]{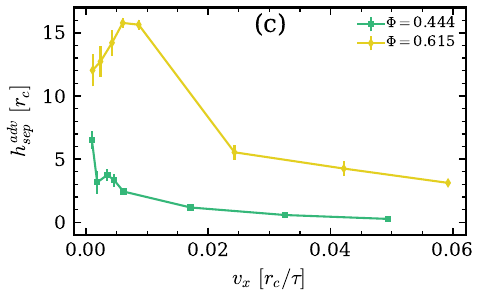}
		\includegraphics[width=8cm]{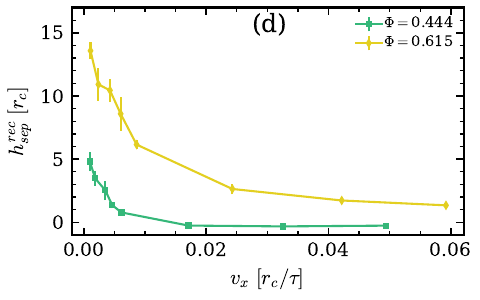}
	\end{centering}
	\caption{(a) Height of the brush at the advancing front $h_{B}^{adv}$ versus droplet velocity. (b) Height of the full ridge at the advancing front $h_{R}^{adv}$ versus droplet velocity. (c) Separation height between the liquid part of the ridge and the brush part at the advancing front $h_{sep}^{adv}$ versus droplet velocity, featuring nonmonotonic behavior at high swelling. (d) Separation height between the liquid part of the ridge and the brush part at the receding end $h_{sep}^{rec}$ versus droplet velocity of the droplet. In (c) and (d) only data for cloaked droplets are shown ($\Phi\geq0.444$) since only those exhibit a separation between the liquid part and the brush part in the ridge.}
	\label{fig:heightvsV}
\end{figure*}

Next we study the influence of the moving droplet on the brush. The droplet pulls up on the brush at the three-phase contact line, due to the interfacial tension. Beyond the cloaking transition, the wetting ridge separates into two ''phases''. One phase still includes the brush chains, and the other is purely composed of free chains. A similar phase separation has also been observed for swollen PDMS gels \citep{jensen2015wetting,cai2021fluid,hauer2023phase}. To quantify the effect of the moving droplet on the brush, we measure how the heights of the brush and the full ridge (brush and free chains) at the three-phase contact line vary with the velocity of the moving droplet. The heights are calculated as the highest points in the contours shown in Figure \ref{fig:contours}, measured from the grafting surface, both for the brush and the full substrate, and both for the receding end and the advancing front. In addition, a quantity of interest that has been studied in experiments is the separation height between the two phases \citep{hauer2023phase}. It is calculated as the difference between the brush height and the height reached by the free chains. Since no separation is observed below the cloaking transition, we can only calculate the separation height for the two densities $\Phi = 0.444$ and $\Phi = 0.615$.

Figure \ref{fig:heightvsV}~a) shows the height of the brush at the advancing front $h_{B}^{adv}$ versus the steady state velocity $v_x$ of the droplet. The brush height at the advancing front always decreases with increasing velocity. The decrease is less pronounced for high swelling. This is due to the fact that the more the brush is swollen by lubricant, the less it is affected by the presence of the droplet, as can be seen in Figure \ref{fig:equilAngles_Ridge}~b) in the Supplementary Information. This observation has implications for the contribution of the brush to the total friction force on the droplet: as the droplet moves, it pulls the brush at the advancing front. The further out the brush is pulled, the larger the viscoelastic dissipation within the brush. 

Figure \ref{fig:heightvsV}~b) shows the height of the full ridge at the advancing front $h_{R}^{adv}$ versus the steady state velocity $v_x$ of the droplet. For low lubrication, the ridge height decreases as the speed of the droplet increases. However, for $\Phi=0.615$ something interesting happens: With increasing droplet velocity, the ridge height first increases up to a maximum at $v_x\approx0.01$, then it starts decreasing again. Figure \ref{fig:heightvsV}~c) shows the separation height at the advancing front $h_{sep}^{adv}$ versus the steady state velocity $v_x$ of the droplet. For $\Phi=0.444$ and within the confidence intervals, the trend follows the expected decrease with velocity, while for $\Phi=0.615$ the trend mirrors that of the ridge height $h_{R}^{adv}$. Finally, Figure \ref{fig:heightvsV}~d) shows the separation height at the receding end $h_{sep}^{rec}$ versus the steady state velocity $v_x$ of the droplet. The separation between the brush and the liquid ridge at the receding end gradually decreases as the velocity of the droplet increases. 

The peculiar behavior at the advancing front could be explained as follows. As the droplet rolls, it drags material from the receding end onto the advancing front. The material is transported as part of the cloak. For all velocities, material is depleted from the receding end, hence the separation height decreases. At the advancing front, however, material will be deposited. At low velocities, the material is deposited there and the ridge grows, leading to an increase in the separation height. This is the case until a threshold velocity is reached. Then the droplet starts rolling over the advancing ridge leading to a decrease in the separation height.


\subsection{Flow Field}

\begin{figure}[!ht]
	\begin{centering}
		\includegraphics[width=8cm]{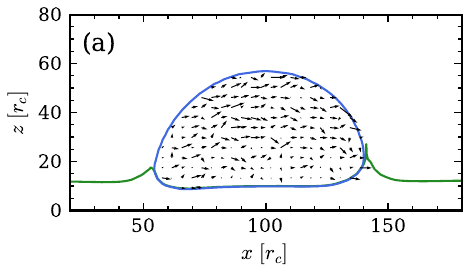}
		\includegraphics[width=8cm]{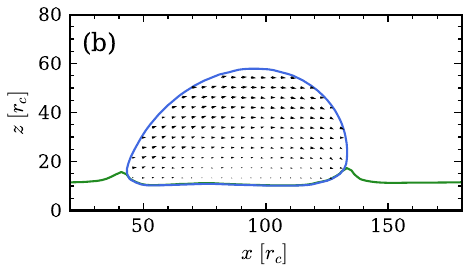}
		\includegraphics[width=8cm]{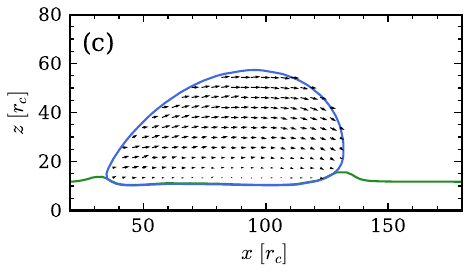}
		\includegraphics[width=8cm]{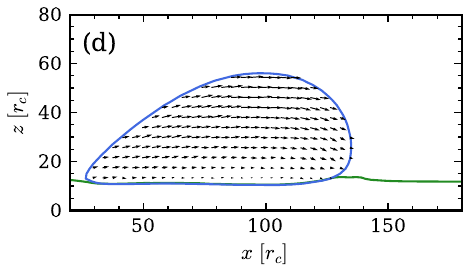}
	\end{centering}
	\caption{Flow field inside the droplet in the laboratory frame for different applied bulk forces and lubricant fraction $\Phi=0.615$. Green contour represents the substrate, blue contour the droplet. (a) $F=8\times10^{-5}k_BT /r_c$, (b) $F=20\times10^{-5}k_BT /r_c$, (c) $F=30\times10^{-5}k_BT/r_c$, (d) $F=40\times10^{-5}k_BT/r_c$. In (b), (c), and (d) the scale of arrow lengths (representing the amplitude of the local flow velocity) is the same, while in (a) the arrow lengths are amplified. The low flow velocity near the substrate indicates little to no slipping.}
	\label{fig:flowFieldLab}
\end{figure}

\begin{figure}[!ht]
	\begin{centering}
		\includegraphics[width=8cm]{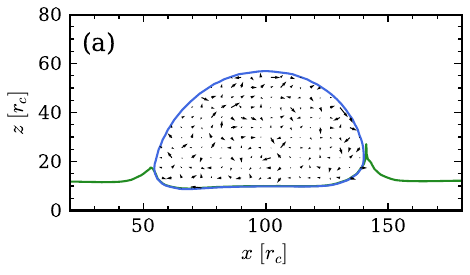}
		\includegraphics[width=8cm]{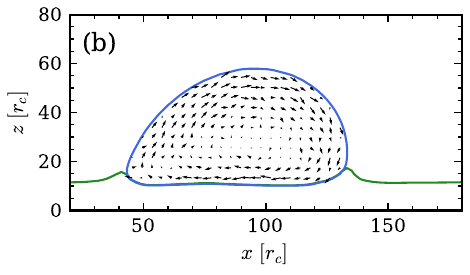}
		\includegraphics[width=8cm]{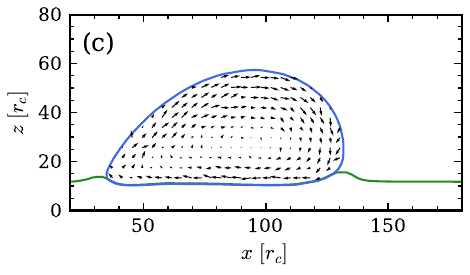}
		\includegraphics[width=8cm]{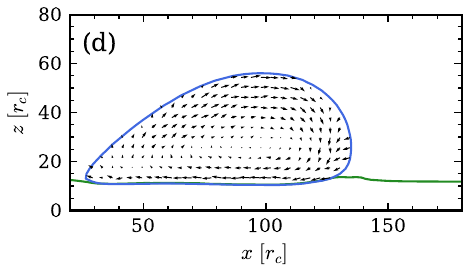}
	\end{centering}
	\caption{Same as Figure \protect\ref{fig:flowFieldLab} in the co-moving frame of
 the center of mass of the droplet. The scales of arrow lengths are different for all figures. The flow field shows clear signs of rolling motion.}
	\label{fig:flowField}
\end{figure}

To understand the origin of the friction, we need to know whether the drop is rolling, sliding, or a combination of both. To answer this we calculate the flow field within the droplet. Figure \ref{fig:flowFieldLab} shows the flow field in the laboratory frame for a selection of forces. \rev{As above, we select a slab of the droplet of thickness $2r_c$ along a plane parallel to the direction of motion and analyze the flow in this slab}. To obtain the final flow fields we calculate a flow field from each simulation by taking an average over 150 snapshots, and subsequently average the flow fields of 5 independent simulations. For $F=8\times10^{-5} k_BT/r_c$ (Figure \ref{fig:flowFieldLab}~a), the average local flow is smaller than the statistical error due to thermal motion within the droplet, hence flow profiles are not discernible. For the other forces $F=20\times10^{-5} k_B T/r_c; \, F= 30\times10^{-5}k_B T/r_c;\, F=40\times10^{-5}k_BT/r_c$ in (b), (c), and (d) respectively, there is a clear net flow toward the right. At the interface between the droplet and the brush, the velocity is small indicating that the droplet rolls with little to no sliding even in the presence of lubricant. This is different from previous simulations of polymer droplets on hard corrugated substrates\cite{servantie2008statics}, where significant slip was observed at the surface.

Figure \ref{fig:flowField} shows the flow field in the co-moving frame of the center of mass of the droplet. Again, extracting a flow field is not possible at the lowest force due to the random thermal motion. For stronger forces, the flow field profiles clearly feature a rolling motion. Interestingly, the flow field inside the droplet seems to feature two vortices, one close to the center of the droplet and one towards the back end. Similar two-vortex structures have been visualized experimentally in flow profiles inside adhering droplets in shear flow~\cite{burgmann2022inner}. Numerical Lattice Boltzmann simulations of rolling droplets on flat and structured hydrophobic surfaces have also revealed complex flow patterns inside the droplets~\cite{abubakar2021droplet}.



\subsection{Cloak Structure and Oil Transport}

\begin{figure}[!t]
	\begin{centering}
		\includegraphics[width=8cm]{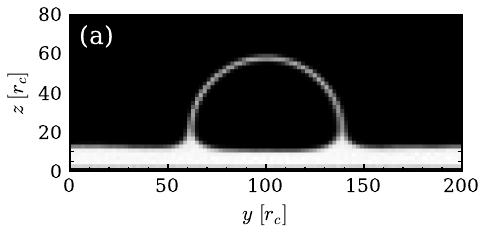}
		\includegraphics[width=8cm]{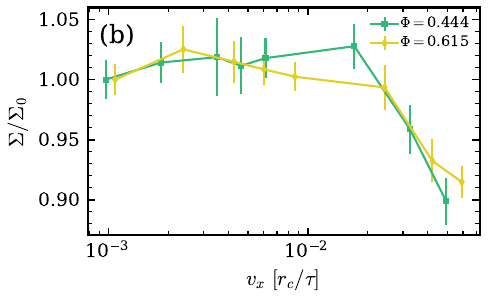}
	\end{centering}
	\caption{(a) Example of a polymer density map in a cross-section perpendicular
 to the droplet velocity at $\Phi = 0.615$.
(b) Normalized Surface density of free chains at the top of the droplet $\Sigma/\Sigma_0$ versus droplet velocity ($\Sigma_0 = 4.0 /r_c^2$ at $\Phi=0.444$, $\Sigma_0 = 4.4/r_c^2$ at $\Phi=0.615$).}
	\label{fig:cloakvsV}
\end{figure}

As real cloaked droplets move on PDMS substrates, they carry the lubricant with them, thus depleting the lubricant and reducing the associated favorable qualities of the substrate. To investigate the effect of the droplet motion on its cloak, we study how the cloak thickness is affected by the droplet velocity. We now section the droplet along a plane perpendicular to the direction of motion. We choose a section centered at the center of mass of the droplet and of thickness $2r_c$ and calculate the local density of polymers $\rho(y,z)$ within that section. Such a density map is shown in Figure \ref{fig:cloakvsV}~a). To measure the variation in the cloak we calculate an effective surface density at the top of the droplet 
\begin{equation}
	\Sigma=\int_{z_1}^{z_2}dz\rho(y=0,z)
\end{equation}

with $z_1$ corresponding to a point within the droplet, and $z_2>z_1$ a point above the cloak. Figure \ref{fig:cloakvsV}~b) shows $\Sigma$ versus the droplet velocity normalized by the value at the lowest velocity $\Sigma_0$. For low velocities the cloak is not affected, but for the highest two velocities the value slightly decreases by about $10\%$.

The fact that the cloaking layer is not strongly affected by the motion, combined with the fact that the area of the droplet increases at larger velocities due to shape deformations, implies that the transport of lubricant is expected to be enhanced at higher velocities.


\section{Summary and Conclusion}

In sum, we have studied the effect of adding lubricant on the motion of droplets on polymer brush coated surfaces under the influence of an external, e.g., gravitational, body force. The simulation model is adapted to describe water droplets on surfaces with PDMS brushes and free PDMS chains acting as lubricant. One characteristic of this system is the existence of a cloaking transition \cite{badr2022cloaking}: beyond a critical lubricant content, the water droplet is fully covered by a thin layer of free PDMS chains. Hence our study also addresses the question how such a cloaking layer affects the droplet motion.

In general, we find that the droplet velocity $v$ increases with the applied force $F$ following a power law, $v \propto F^{1/\alpha}$. With increasing velocity, the droplet deforms and elongates. The receding contact angle decreases roughly linearly as a function of $v$, while the advancing angle remains roughly constant.

Within the investigated amounts, lubrication has a surprisingly small effect on this behavior. Adding a limited amount of lubricant chains only slightly reduces the friction force between the droplet and the surface. This is already suggested indirectly by the experiments, where it was found that the friction does not depend on the method used to prepare the polymer brushes. In particular, whether or not free polymers have been washed out from the brushes after synthesis did not seem to make any difference. However, the degree of swelling could not be determined in the experiments. In the simulations, the velocity of droplets subject to a given driving force does not change at all (within the statistical error) if one adds lubricant amounts below the cloaking transition. Above the cloaking transition, the velocity somewhat increases, i.e., the friction is somewhat reduced, but by less than a factor 3/4 at the highest lubricant content. The power law exponent $\alpha$ increases slightly for cloaked droplets, but by less than 20\%.

Lubricants also have no discernible effect on the macroscopic properties of the droplet, i.e., its shape as a function of velocity, and the advancing and receding contact angles. They do however affect the local structure at the contact line, i.e., the wetting ridge. In particular, the ridge height of strongly cloaked droplets exhibits a maximum as a function of droplet velocity, suggesting a transition between a low-velocity regime where the droplet pushes the wetting ridge in front of it, to a regime where it partially rolls over it.

The reason for the unexpected low lubricating effect of free chains in the brush becomes clear when analyzing the pattern of the flow field inside the droplet. The friction results from an interplay of energy dissipation at the adhesive surface, at the contact line (the wetting ridge), and inside the droplet. In our system, we find that the flow velocity is very small at the interface between droplet and substrate, thus the droplet does not slide, and the dissipation at the adhesive surface can be neglected. The dissipation at the wetting ridge does contribute - as can be inferred from the fact that the friction force is affected by the cloaking transition, but this contribution seems to be comparatively small. \rev{We then argue that} the dominating contribution to dissipation is the shear flow inside the droplet, which is not affected by the lubricant. For droplets of fixed shape with fixed flow patterns, one would expect the resulting friction force to scale linearly with the droplet velocity\cite{servantie2008statics}. However, the droplets can partly reduce the dissipation at high velocities by changing their shape, which affects the flow profiles as discussed in Section ''Flow Field''. As a result, the increase of the friction force with droplet velocity is only sublinear.

The question remains why the power exponent $\alpha$ measured in the simulations ($\alpha \sim 0.6-0.7$) differs from the one measured in the experiments ($\alpha \sim 0.25$). Although we have studied large droplets containing almost a million particles, they are still
microscopic compared to the droplets considered in the experiments, which have sizes of the orders of millimeters. Nevertheless, the 
Reynolds numbers in the simulations and experiments are comparable ($\text{Re} = \frac{\rho v R}{\mu} \approx 1$ in the simulations for velocity $v=0.04 r_c/\tau$ and droplet radius $R \sim 40 r_c$, and $\text{Re} \sim 1$ in the experiments for $v = 1$mm/s and $R \approx 1$mm), suggesting that the flow fields may also be comparable. On the other hand, the experimental drop sizes are comparable to the capillary length such that gravity may become important. The influence of droplet size and gravity on the friction force will be an interesting subject of future studies.

We should also note that, in the present study, we have only studied lubricant fractions where the brush is not yet fully saturated. If
we had much more lubricant, it will form a thick film above the brush, and the situation will change. The droplet will then be immersed in the lubricant film and the shear flow inside the film will significantly contribute to the dissipation.
 

\section{Acknowledgments}
This work was funded by the German Science Foundation (DFG) within the priority program 
SPP 2171 (Grant No. 422796905, projects Schm 985/22 and VO 639/16). Further support is acknowledged from the DFG-funded Graduate School
RTG 2516 (Grant No. 405552959): RGBM and LH are
associated members, FS is a member. 
The simulations were partly carried out
on the supercomputer system Mogon NHR S\" ud-West at Johannes Gutenberg University Mainz.
RGMB thanks Leonid Klushin for useful discussions.



\bigskip



\bibliography{refs.bib}

\clearpage

\renewcommand\thefigure{S.\arabic{figure}}
\setcounter{figure}{0}
\renewcommand\theequation{S.\arabic{equation}}
\setcounter{equation}{0}

\section{Supplementary Information}

\subsection{Characteristics of equilibrium droplets}

\begin{figure}[!htb]
	\begin{centering}
		\includegraphics[width=8cm]{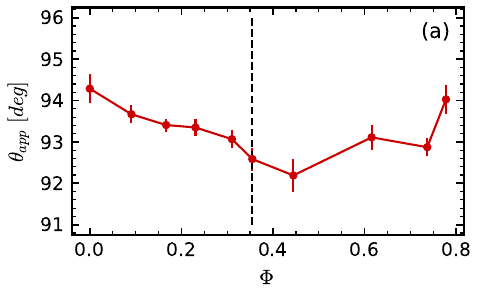}
		\includegraphics[width=8cm]{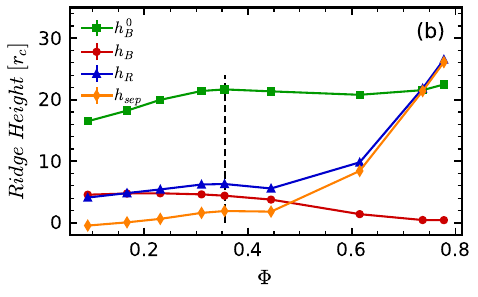}
	\end{centering}
	\caption{Characteristics of equilibrium droplets on lubricated brushes as a function of lubricant fraction $\Phi$ (after Reference \citenum{badr2022cloaking}). The vertical dashed line indicated the onset of the cloaking transition. (a) Apparent contact angle. 
 (b) Height of the brush ($h_B$, red) and the full ridge ($h_R$, blue) relative to the
 height of the unperturbed brush, difference between the two ($h_{sep}$, orange), and absolute height of the brush ($h_B^0$, green) at the three phase contact line.
 }
	\label{fig:equilAngles_Ridge}
\end{figure}

The surface tensions, the contact angle, and the properties of the wetting ridge in equilibrium droplets have been determined in previous work\cite{badr2022cloaking}. For the convenience of the reader, we reproduce some pertinent results in Figure \ref{fig:equilAngles_Ridge}. Figure \ref{fig:equilAngles_Ridge} a) shows the macroscopic contact angle as a function of lubricant fraction $\Phi$ and demonstrates that it stays almost constant. Figure \ref{fig:equilAngles_Ridge} b) characterizes the wetting ridge for varying $\Phi$ in terms of the heights of the brush $h_B$ and the full ridge (brush and lubricant, $h_R$) at the three phase contact line, taking the unperturbed brush as a baseline. The curves for $h_R$ and $h_{sep} = h_R - h_B$ demonstrate that the wetting ridge grows only slowly with increasing $\Phi$ below the cloaking transition, but rises rapidly above the cloaking transition. Also shown in this figure is the absolute height of the brush ($h_B^0$), measured with respect to the grafting surface. This quantity grows below the cloaking transition and
then saturates at a value corresponding to the thickness of an unperturbed saturated brush ($\sim 21 r_c$) above the cloaking transition.

\subsection{Poiseuille flow}

\begin{figure}[!htb]
	\begin{centering}
		\includegraphics[width=5cm]{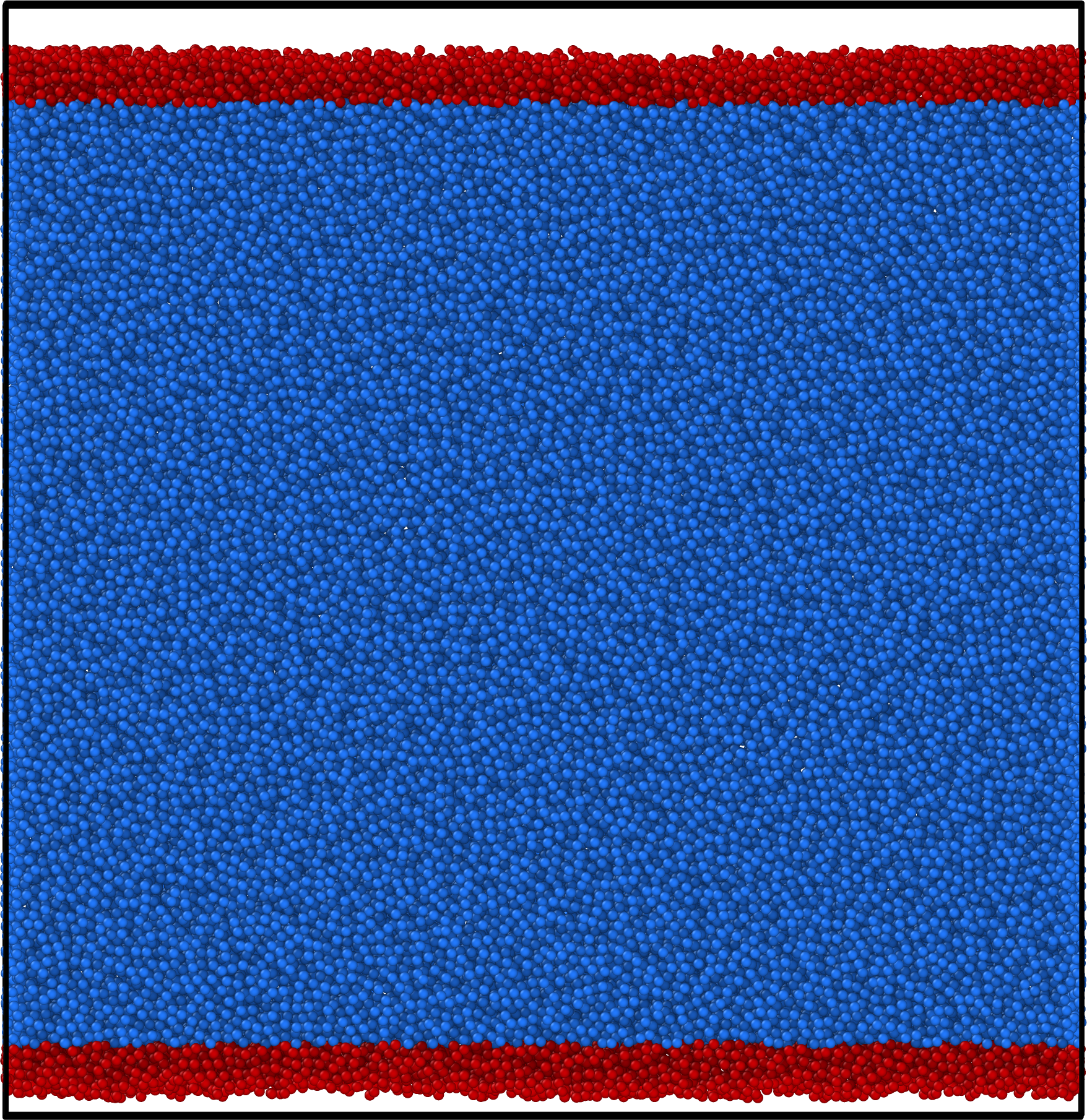}
	\end{centering}
	\caption{Snapshot from the Poiseuille flow simulations. Red particles are fixed while the blue particles are allowed to flow.}
	\label{fig:poiseuilleSnapshot}
\end{figure}

In order to get an estimate of the viscosity, we set up Poiseuille flow simulations. We start with a liquid slab at equilibrium, fix some particles at the top and bottom (red particles in Figure \ref{fig:poiseuilleSnapshot}), and apply a constant force per particle to the other particles (blue particles in Figure \ref{fig:poiseuilleSnapshot}). The resulting flow profile has the form

\begin{equation}
	v_x(z)=-\frac{F \rho_w}{2\mu}(z^2-z_0^2)
	\label{eq:poiseuille}
\end{equation}

with $\mu$ the dynamic viscosity, where $F$ is the force per particle, $\rho_w=4$ the number density and $v_x(\pm z_0)=0$.

\begin{figure}[!htb]
	\begin{centering}
		\includegraphics[width=8cm]{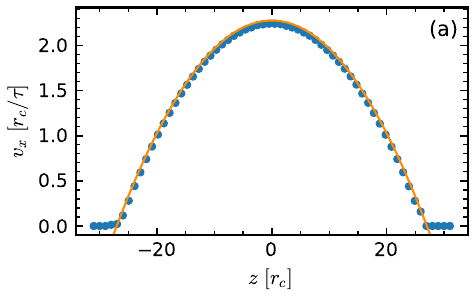}
		\includegraphics[width=8cm]{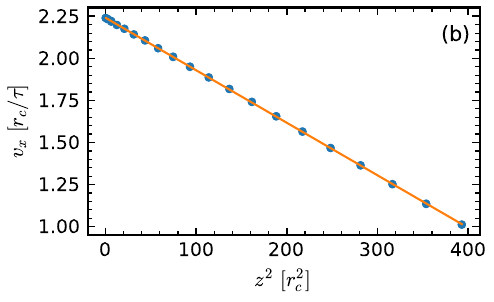}
	\end{centering}
	\caption{(a) Poiseuille flow obtained for a force $F=0.01 k_BT/r_c$ per particle. (b) the velocity profile versus $z^2$ near the middle of the slab. Solid line is a linear fit. The fit gives a value for the dynamic viscosity $\mu = 6.404\pm 0.006~k_BT r_c^{-3} \tau$. The error is propagated from the standard error on the slope. The solid line in the top figure is an evaluation of Eq. \ref{eq:poiseuille} using the calculated viscosity and $z_0=27$.}
	\label{fig:poiseuilleFlow}
\end{figure}

The flow for a force $F=0.01$ is shown in Figure \ref{fig:poiseuilleFlow} a). The flow is parabolic as expected. To calculate the viscosity, we take velocity values for $z\in[-20,20]$ and use linear fitting to determine the parameters $B$ and $A$ of $v_x=B z^2 + A$ (see Figure \ref{fig:poiseuilleFlow} (b) ). Then from the slope $B$ we calculate the dynamic viscosity $\mu=\frac{F\rho_w}{2B}$. For the data shown in Figure \ref{fig:poiseuilleFlow} we obtain $\mu = 6.404\pm 0.006~k_BT r_c^{-3} \tau$.

\end{document}